# Malware Classification Using Deep Boosted Learning

Muhammad Asam[1, 2], Saddam Hussain Khan[1, 2], Tauseef Jamal[4], Umme Zahoora[1, 2], Asifullah Khan[1, 2, 3*]

[1]Pattern Recognition Lab, Department of Computer & Information Sciences, PIEAS, Islamabad 45650, Pakistan

[2]PIEAS Artificial Intelligence Center (PAIC), PIEAS, Islamabad 45650, Pakistan

[3]Center for Mathematical Sciences, PIEAS, Nilore, Islamabad 45650, Pakistan

[4]Department of Computer and Information Sciences (DCIS), PIEAS, Nilore, Islamabad 45650, Pakistan

asif@pieas.edu.pk

*Abstract*

Malicious activities in cyberspace have gone further than simply hacking machines and spreading viruses. It has become a challenge for a nations' survival and hence has evolved to cyber warfare. Malware is a key component of cyber-crime, and its analysis is the first line of defence against attack. This work proposes a novel deep boosted hybrid learning-based malware classification framework and named as Deep boosted Feature Space-based Malware classification (DFS-MC). In the proposed framework, the discrimination power is enhanced by fusing the feature spaces of the best performing customized CNN architectures models and its discrimination by an SVM for classification. The discrimination capacity of the proposed classification framework is assessed by comparing it against the standard customized CNNs. The customized CNN models are implemented in two ways: softmax classifier and deep hybrid learning-based malware classification. In the hybrid learning, Deep features are extracted from customized CNN architectures and fed into the conventional machine learning classifier to improve the classification performance. We also introduced the concept of transfer learning in a customized CNN architecture based malware classification framework through fine-tuning. The performance of the proposed malware classification approaches are validated on the **MalImg** malware dataset using the hold-out cross-validation technique. Experimental comparisons were conducted by employing innovative, customized CNN, trained from scratch and fine-tuning the customized CNN using transfer learning. The proposed classification framework DFS-MC showed improved results, Accuracy: 98.61%, F-score: 0.96, Precision: 0.96, and Recall: 0.96.

# 1. INTRODUCTION

Software designed in a malicious commitment to harm user or system, come under the category of malware. Malware can harm a system to any level of damage. It ranges from gaining system access, deleting file, ransom demand or even complete sabotage of system. All this happens in backend, without user knowledge. An adequate increase has been found in credential harvesting using malware and well-established tactics in the recent past. Even in the COVID-19 pandemic traverse, Microsoft reported 16 different state-level actors who targeted commercial and academic institutions for stealing vaccine-related research knowledge. These threat actors have rapidly gained sophistication over the past years. They are skilled, persistent and launches attacks, which are harder to spot. AV-TEST Institute reports more than one billion infected files in January 2021 only [1]. These malware files are morphed into different combinations and variations for the sake of evading detection. Anti-malware techniques are employed for protection against malware invasion. In this regard, categorization of malware is necessary for the precise anti-malware solution [2].

Customarily, the challenge of malware classification is addressed using static and dynamic techniques. The static malware classification approach tries to find out the known signatures in the malware file. In this approach, malware is identified by a sequence of bytes known as malware signature. File hashes are also used for this purpose [3]. This technique results in a lower false positive (FP) rate [4]. However, these signatures can be altered to evade detection from malware scanners. Signature- malware detection is fast and effective in detecting malware whose signatures are known. Static malware detection technique is incapable of recognizing newly released malware [5]. Malware obfuscation is another challenge faced by the static malware classification and detection technique. This malware obfuscation technique includes code manipulation, instruction substitution, register reassignment and dead-code insertion [6].

Dynamic malware classification examines malicious activities and footprints during execution [3]. This execution is performed in a controlled environment. This technique is proficient to detect obfuscated and new malwares. As dynamic techniques look for the anomalies in behaviours of software at runtime, so they can produce large false-positives (FP). Time and excess resource utilization are another drawbacks of behaviour-based classification [7]. Traditional anti-malware techniques are unable to detect complex variants and new attacks. However, classification using image-based malware data has shown to be very promising [8].

Machine Learning (ML) evolution has achieved breakthrough accomplishments in terms of accuracy and scalability in a variety of fields [9]. This evolution played a radical role in shifting from traditional to AI-powered anti-malware strategies. Deep Learning (DL), a sub-field of ML, can even solve more complex problems. DL methodologies learn from experience hierarchy. This ability of DL is due to the learning of data representations at multiple deep feature extraction stages [10]. DL algorithms discover representations for target class in terms of input data distribution [11]. It is performed at multiple levels. Higher-level learns features that are extracted from lower level features. This hierarchical learning makes the computer acquire the complicated concept of the problem by understanding the simpler one. Also, DL is very prospective in image-based classification problems.

ML-based intensive efforts have been utilized for malware analysis. Convolutional Neural Network (CNN) has been successfully employed to classify the malware families [12]. In this work, customized CNNs are used in an end-to-end manner and as a feature extractor with an ML classifier. The feature hierarchies learned from customized CNNs are assigned to SVM for malware classification. Finally, deep features extracted from customized transfer learning (TL)-based fine-tuned CNN (ResNet-18 and DenseNet-201) architectures are concatenated and fed into the traditional ML classifier. The significant contributions of our research work are below:

1. The idea of residual learning and the concept of blocks in CNN are exploited to learn the effective representation. Additionally, we enhanced the feature space diversity by combining them to improve the malware family segregation.
2. The classification power is enhanced by integrating the principle of structural and empirical risk minimisation.
3. The performance of the novel malware classification framework is analysed with several CNN architectures showing high classification performance while significantly decreasing the number of false-positives and false-negatives.

The rest of the paper is organized as follows. The next section highlights the related work in the field of malware classification. Section 3 explains our novel classification framework methodology, while section 4 discusses the experimental setup. Section 5 presents the result analysis and discussion of our work. Section 6 concludes the paper.

## 2. RELATED WORK

Static and dynamic malware analysis techniques have been applied extensively for classification and detection problem. The evolution of machine learning has opened many horizons for analysis and prediction for both malware analysis techniques. Visual malware image-based classification is an addition in the field of malware analysis. In 2011, Nataraj et al. [13] identified the texture features in the malware visual image file. These files are obtained by interpreting the byte code of portable executable (PE) binary file to the grey level of the pixel value in the image. They extracted the texture features from the malware image using wavelet decomposition. Machine learning technique of K-nearest neighbour is applied on these features afterward. This technique achieved 98% Accuracy on the benchmark malware dataset consisting of 25 malware families. Wavelet transformation and support vector machine (SVM) is applied for feature extraction and training the artificial neural network in [14]. A novel lightweight malware classification technique for the Internet-of-Things is focussed in [15]. Lightweight CNN is applied on one channel binaries collected from the IoT network. This proposed approach reached the accuracy of 94.0% good ware and DDoS malware. An improved CNN based model for multi-family classification is presented in [16]. A hierarchical ensemble neural network is proposed to exploit the trace detection in Intel processors [17]. These traces are converted into time series data of images. A deep CNN is trained on the image data. Sang Ni et al. extracted malicious op-code sequences and converted them to images [18]. Their classification achieved Accuracy of 98.86% using CNN.

Virtual data of the malware attacks is generated using Generative Adversarial Networks (GANs) [19]. It helped in achieving the capability of Zero-day malware attack detection. It is reported to be 95.74% accurate. Q Le et al. proposed a deep learning approach using LSTM and CNN [20]. It achieved Accuracy of 98.8% along with improved time efficiency. Android systems are not far away from malware attacks. Another malware detection framework is proposed for the android environment using neural networks [21]. A static malware detection approach is applied on PE file using malware benchmark data, EMBER, reporting the classification efficiency 98.9% [22]. Local-Global Malicious Pattern (LGMP) is presented in [23]. It used hybrid visual features of binary files for malware detection. An ensemble of CNN is proposed in [24].

# 3. METHODOLOGY

Malware categorization is a complex-mapping problem after malware identification. Classification of malware using a single original feature representation is challenging because of diverse properties of different malwares. We have introduced a new classification framework, based on deep feature learning and classical ML techniques for automatic discrimination of malware families into 25 classes[dataset]. The proposed malware classification framework can be viewed in two main phases. In the first phase, data augmentation is performed. While in the second phase, four different (Classification schemes) deep CNN based techniques are employed for malware analysis. These schemes are (i) implementation of existing CNN using training from scratch, (ii) incorporation of TL concept by fine-tuning malware classification models, (iii) deep feature-based Malware classification and (iv) the proposed approach Deep Feature Space-based Malware Classification (DFS-MC).

In the proposed DFS-MC approach, ResNet-18 and DensNet-201 are customized and fine-tuned according to our problem space. Transfer learning based training is used for these networks. Deep ensemble features are extracted from these customized CNN models. Finally, the malware classification is achieved by using SVM on the deep ensemble feature space. The overall setup of the proposed classification framework is shown in Figure 1. Our proposed framework is stepwise explained in the following.

   Malware classification framework
   1. The Proposed Approach DFS-MC
   2. Implementation of well-established CNN architectures
       2.1. Deep feature-based classification
       2.2. Softmax probabilistic-based classification

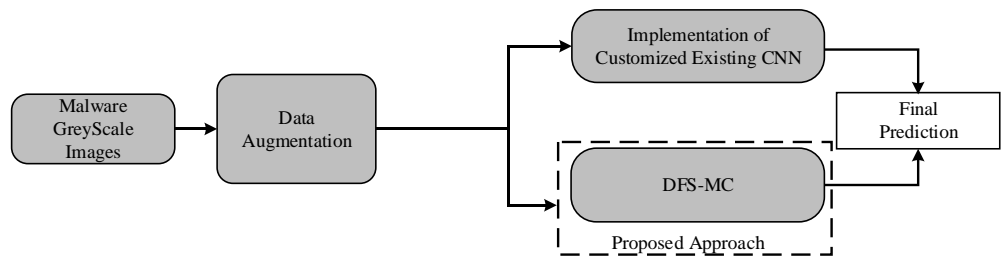

**Figure 1.** Brief overview of the proposed malware classification framework.

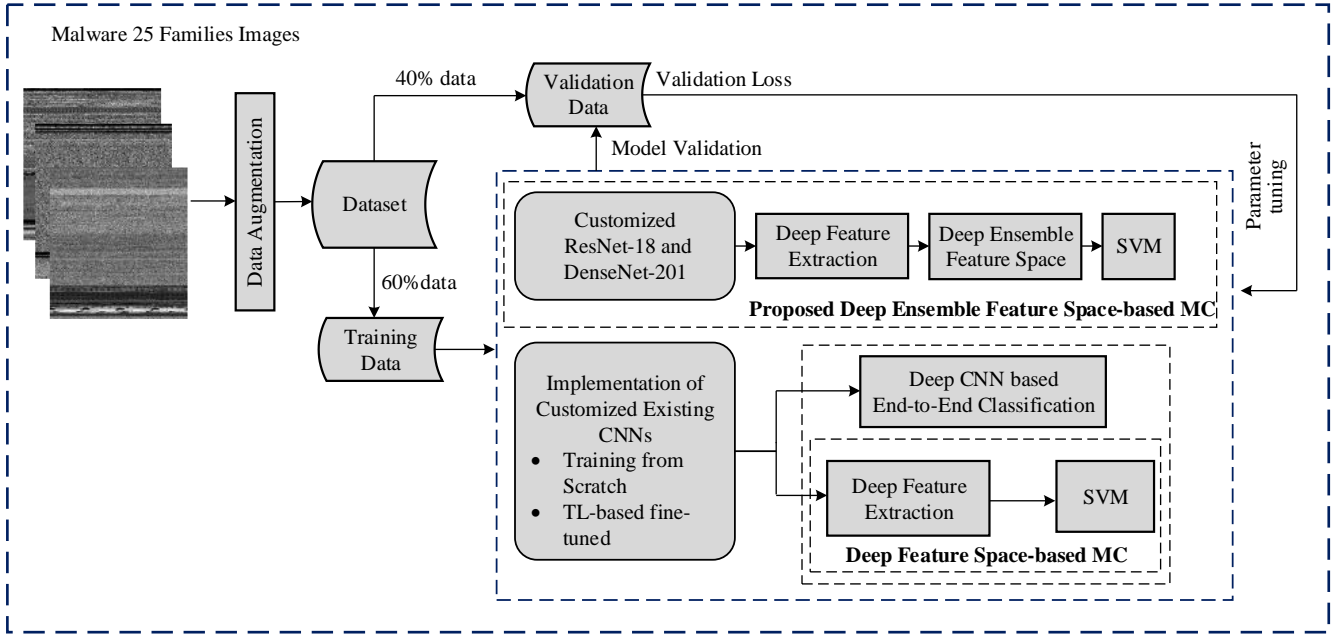

**Figure 2.** Training Phase of malware classification approaches.

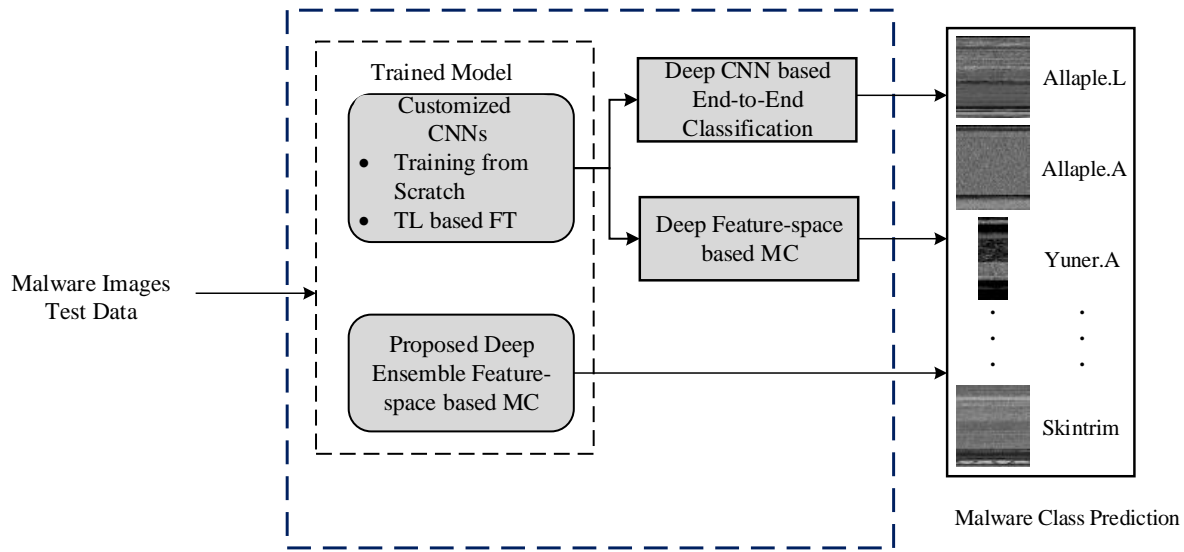

**Figure 3.** Testing phase of malware classification approaches.

## 3.1. Dataset partitioning

In this work, we employed a 60-40 dataset partitioning scheme for training and testing phases. Normally 80-20, 75-25 or sometimes 70-30 dataset partitioning is present in the literature. In our case, the malware-based dataset is limited. This scheme is used to make our model more robust to the unseen malware data. Dimensions of the dataset matters in this regards.

**Table 1.** MalImg dataset description and class instance details.

| No. | Family | Family Name | No. of Instances |
|---|---|---|---|
| 1 | Worm | Allaple.L | 1591 |
| 2 | Worm | Allaple.A | 2949 |
| 3 | Worm | Yuner.A | 800 |
| 4 | PWS | Lolyda.AA 1 | 213 |
| 5 | PWS | Lolyda.AA 2 | 184 |
| 6 | PWS | Lolyda.AA 3 | 123 |
| 7 | Trojan | C2Lop.P | 146 |
| 8 | Trojan | C2Lop.gen!g | 200 |
| 9 | Dialer | Instantaccess | 431 |
| 10 | TDownloader | Swizzot.gen!I | 132 |
| 11 | TDownloader | Swizzor.gen!E | 128 |
| 12 | Worm | VB.AT | 408 |
| 13 | Rogue | Fakerean | 381 |
| 14 | Trojan | Alueron.gen!J | 198 |
| 15 | Trojan | Malex.gen!J | 136 |
| 16 | PWS | Lolyda.AT | 159 |
| 17 | Dialer | Adialer.C | 125 |
| 18 | TrojanDownloader | Wintrim.BX | 97 |
| 19 | Dialer | Dialplatform.B | 177 |
| 20 | TrojanDownloader | Dontovo.A | 162 |
| 21 | TrojanDownloader | Obfuscator.AD | 142 |
| 22 | Backdoor | Agent.FYI | 116 |
| 23 | Worm:AutoIT | Autorun.K | 106 |
| 24 | Backdoor | Rbot!gen | 158 |
| 25 | Trojan | Skintrim.N | 80 |
| Total | | | 9342 |

## 3.2. *Data augmentation*

Deep learning models overfit an insufficient amount of data. Therefore, a considerable amount of data is required for practical training and achieving good generalization. Data augmentation refers to augmenting the base data to increase data samples [25], [26]. In this work, we have augmented the training dataset to improve the generalization and make the proposed classification framework robust against various malware data. It thus makes the proposed classification framework effective for classifying malware families. The implemented augmentation strategy includes several transformations, such as reflections, scaling, rotation, and shear, as shown in Table 1.

Training samples were initially augmented to improve the model generalization. Our proposed framework is trained using these augmented samples.

**Table 2.** Dataset augmentation details

| Augmentation | Parameters |
|---|---|
| Rotate | [0, 360] degrees |
| Shear | [-0.05, 0.05] |
| Reflection | X: [-1, 1], Y: [-1, 1] |
| Scale | [0.5, 1] |

### *3.3. The Proposed the Deep Feature Space-based Malware Classification (DFS-MC)*

The issue of malware classification is addressed by proposing a hybrid learning scheme consisting of a deep feature space of CNN and SVM from classical ML techniques as a robust classifier. This new technique is called Deep Feature Space-based Malware Classification (DFS-MC). We exploited the potential of residual and block-based learning to generate prominent features from malware visual images. Whereas, softmax layer (Eq. 7) was replaced with an SVM classifier to increase the generalisation ability. The performance of well-established CNN architectures is analysed by using feature extraction and employing SVM as a classifier. Also, the performance is quantised using softmax probabilistic classification. Transfer learning based training and training from scratch is also used. Two architectures are selected based on their classification performance. These two architectures are described below.

**ResNet-18:**- Deep learning experts started building deeper networks for the sake of a better understanding of the complex features. However, the accuracy of the deeper networks was found saturated or even degraded in some cases. During the training of the networks, errors are calculated, and the gradient is determined. This gradient is backpropagated to the initial layers for weight update. This gradient becomes weaker and weaker as it reaches the initial layers of the deeper network. The weights at these layers are not updated or slightly updated, resulting in no learning at these lower layers effectively. Residual Network (ResNet-18) solved this problem by introducing skip connections, also known as identity connection [27]. Skip connections help the gradient to flow through an alternate path and allow the model to learn identity hypothesis function. They provide a mean for information flow from earlier to later layers of the model. ResNet-18 uses a bottleneck residual block design to increase the performance of the network. ResNet-18 is 18 layers deep.

$$\mathbf{x}_{s,t} = \sum_{a=1}^{m} \sum_{b=1}^{n} \mathbf{x}_{s+a-1,t+b-1} \mathbf{w}_{a,b} \qquad (1)$$

$$y = f(x, \{w_i\}) + x \qquad (2)$$
$$y = f(x, \{w_i\}) + w_s x \qquad (3)$$

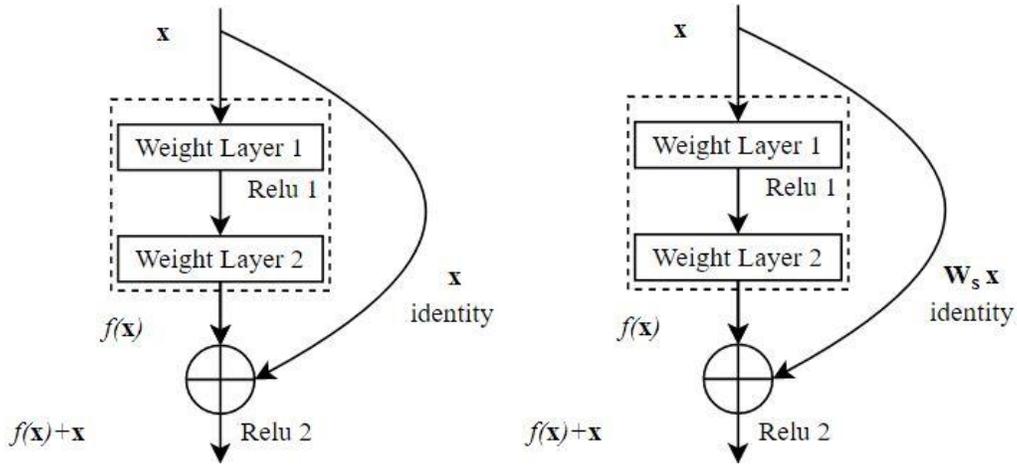

**Figure 4.** Skip Connection Building Block

In Eq. (1), the convolutional operation is employed (assuming the filter is symmetric). Input feature map, of size S x T, is represented by **x**, where kernel, of size *m* x *n*, is represented by **w**. The resultant feature map is represented by **x**, where *s* and *t* run from 1 to S - *m* + 1 and T - *n* + 1, respectively, as depicted in Eq. (1).

Eq. (2) and Eq. (3) describe the residual operation, where $w_i$ and $w_s$ represents the weight layers of 3x3 and 1x1(skip connection) convolutional operation, respectively.

**DenseNet-201:-** This architecture connects each layer in a feed-forward manner. Other architectures with L layer deep have L connections, while this architecture has L*(L+1)/2 connections. Features of the proceeding layers are used as input to the current layer, and features extracted in the current layer is used as input to all subsequent layers. This network is closer to ResNet-18 as inputs are concatenated instead of addition. This small change has a substantially improved vanishing-gradient problem, strengthened feature propagation, encouraged feature reusability and reduced the number of parameters [28]. DenseNet-201 follows simple connectivity rules. This help to integrate the identity mapping and deep supervision properties. The internal representation of this network is very compact. Feature reusability decreased the effect of feature redundancy. So is the reason that DenseNet-201 emerged to be a fine feature extractor for computer vision problems. This network has 201 layers deep.

$$x^l = H_l(x^0|x^1|x^2,...,x^{l-1}) \quad (4)$$

$H_l(.)$ is the non-linear transformation, and $(x^0|x^1|x^2,...,x^{l-1})$ is concatenated feature space up to layer $l$.

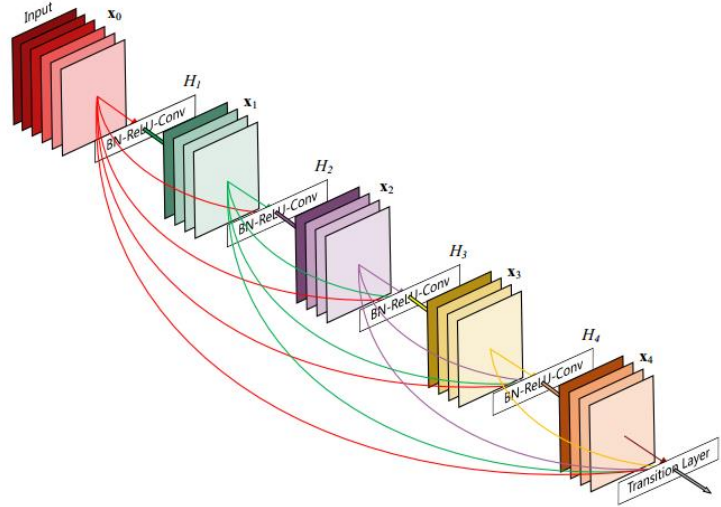

**Figure 5.** Densely connected network architecture [28]

The above findings of both the network architectures are highly appreciated in the form of experimental results that we performed against other architectures. The feature spaces extracted from both the best performing ResNet-18 and DenseNet-201 was concatenated to make a single enriched feature space (Eq. 4).

$$\sigma(v)_i = \frac{e^{V_i}}{\sum_{j=1}^{K} e^{V_j}} \quad (5)$$

$$x_{hybrid} = f_b(f_{ResNet}(x) | f_{DensNet}(x)) \quad (6)$$

$$w^T x + b = 0 \quad (7)$$

$$\min_{w,\xi_i} C \sum_i^N \zeta_i + \frac{1}{2}\|w\|^2 \quad (8)$$

In (Eq. 7), σ represents the softmax activation function, whereas **v** are features from the last layer of customized CNNs and input to softmax. The exponential function for input and normalization vectors are shown, respectively, by $e^{V_i}$ and $e^{V_j}$, and K represents the number of classes. In (Eq. 5-7), $x_{hybrid}$ shows the ensemble feature-space that are generated by concatenation of deep feature space of $f_{ResNet}(x)$ and $f_{DenseNet}(x)$. In (Eq. 7), **x** is an ensemble feature space, whereas $w^T$ is a weight vector orthogonal to hyperplane and **b** is a bias. SVM uses (Eq. 8) to construct an optimal hyper-plane by minimizing the misclassification rate and maximizing the margin between the samples. ζ represents the misclassified samples, while *C* is cost of misclassification that establishes the trade-off misclassification rate and the model's generalization.

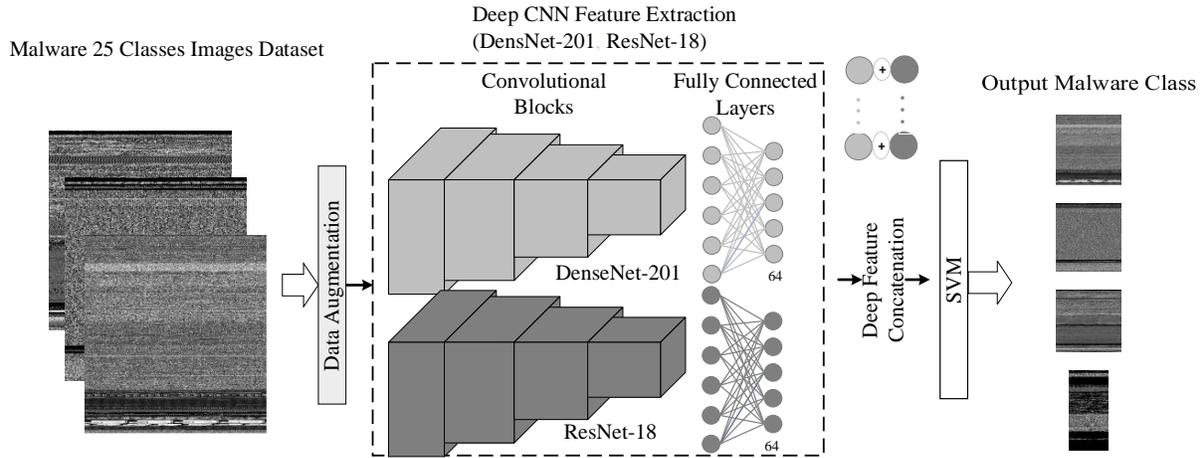

**Figure 6.** Flow diagram of the proposed DFS-MC malware classification approach.

### 3.4. Implementation of CNN architectures

#### 3.4.1. Training Schemes

After the CNN model selection, is the process of training the model according to the problem domain. Training the CNN model is the only way for adjusting weights and biasness of the model. Weights are adjusted by the help of optimization algorithms and backpropagation using training dataset. An ideal training scheme exactly produces the output from the input dataset. We used two main strategies for training the CNNs.

***Training from Scratch:*** We customized different standard well-known deep CNN models according to the compatibility of our proposed malware challenge. These customized architecture are DenseNet-201, GoogleNet, InceptionV3, Xception, Resnet-18, ResNet-50, AlexNet, VGG16 and VGG19 [10], [29]–[31]. Initial and final layers of networks are customized according to the compatibility of the input data set and the target multi-class (25-classes). These CNNs have been extensively used for a wide range of image classification problems and have been used by several researchers for malware family classification [32]. These models vary in block design and architecture. Still, all of them either exploited a single type of pooling operation down the network or replaced the pooling operation with a strided convolution operation for complexity regulation. These architectures are optimized for malware images by training them from scratch on malware dataset by randomly initializing weights from a uniform distribution.

**Transfer Learning:** The core of deep learning is the labelled data repository upon which the model is effectively trained. In case the labelled data in the target domain is limited, the concept of data augmentation and transfer learning is useful [33]. CNNs are parameter hungry and require a large amount of data for training. TL is a technique that has shown promising results for CNN models in the non-availability of a large dataset. It allows the reuse of the already trained models' weight space and prevents highly parameterized models from overfitting by providing a good initial set of weights. It has shown enhanced performance in the field of malware classification challenges [34]. Pre-trained CNN models of the source domain (Image Classification, ImageNet) are adopted for the target domain (Malware Classification, MalImg) with the help of fine-tuning. This fine-tuning is carried out with the help of additional layers or dissecting the existing layers accordingly. This concept is known as domain adaption [34], [35]. We have introduced the concept of TL-Based Fine-Tuning in the customized existing deep CNN models. Transfer learning uses the knowledge of existing domain for a new target domain. We have chosen the two best CNNs among the customized existing ones, which are ResNet-18 and DenseNet-201. We initialized the weights of the ResNet-18 and DenseNet-201 from the pre-trained model's parameter space. Likewise, for a fair comparison, we adapted the same training strategy for the standard state-of-the-art CNNs.

### 3.4.2. Classification Schemes

The following two schemes are applied for the classification of the malware into a particular malware family.

**Softmax probabilistic-based classification: -** Classification of the malware is also performed using the CNNs in end-to-end format. The activation function in the final layer of the neural network decides the class of the input data. The softmax probabilistic function is also used in the decision (final) layer of neural networks for multi-classification problems. This function accepts the input vectors in the form of real numbers and normalizes the inputs to a probability distribution. This output is proportional to the exponentials of its input numbers. Some of the inputs could be negative or greater than 1. After applying this function, each element will be in the range of 0 to 1. This function makes sure that the sum of all our output probabilities is equal to one. This way, they can be interpreted as a probability distribution. In this case, if we want to increase the likelihood of one class, the other has to decrease by an equal amount. The details of the CNN layers are given in Table 3.

**Table 3.** Depth details of deep CNN models

| Models | Depth (Convolutional +Fully-connected Layers) |
|---|---|
| AlexNet | 8 (5+3) |
| VGG16 | 16 (13+3) |
| VGG19 | 19 (16+3) |
| ResNet18 | 22 (20+2) |
| ResNet50 | 55 (53+2) |
| Google Net | 59 (57+2) |
| Xception | 75 (74+2) |
| Inception | 96 (94+2) |
| DesNet-201 | 203(201+2) |

**Deep feature-based ML classification: -** We implemented these CNNs both for softmax probability-based classification and deep feature learning for ML-based classification. The classification ability of the proposed classification framework is enhanced by combining the benefits of both empirical and structural risk minimization. CNNs are high capacity learning models and follow the principles of empirical risk minimization learning theory, which focuses on minimizing training loss. This, sometimes, may lead to overfitting. Contrary to this, a classical ML classifier like SVM is based on a structural risk minimization principle that improves generalization by looking at a test error [36]. Deep feature spaces were extracted from the second last fully connected layer (FC) of these CNNs and were provided to SVM for classification. The feature extraction layer and its respective dimensions are given in Table 4.

**Table 4.** Deep feature extraction layer and its feature matrix dimension for each CNN model

| Pre-trained | | Trained from scratch | |
|---|---|---|---|
| Feature Layer | Feature Dim. | Feature Layer | Feature Dim. |
| Last fc | 64x25 | New fc | 64x25 |

## 4. RESULTS AND DISCUSSION

In order to make our proposed malware classification framework more robust, we implemented the architectures in the data distribution of 60-40 ratio. These experiments are performed using (i) softmax probability-based Malware classification, (ii) deep feature and ML-based Malware classification and (iii) our proposed malware classification approach (DFS-MC). These malware classification models are implemented using training from scratch for MalImg data set in the first place and then using pre-trained models for ImageNet. Performance measures like accuracy and F1-score of training from scratch and TL-based implementation are recorded.

As the first part of our experiments, the customized malware classification models are implemented using both training from scratch, and TL-based using softmax probability-based malware classification. Table 5 shows the results.

It is evident from these experiments that the TL-based approach outperformed the training from scratch implementation. This is because the pre-trained architectures have already been trained on a variety of data. Hence gave good results upon fine-tuning according to our challenge. In the second part, the deep features extracted from CNN architectures are fed to ML for malware classification. Results are shown in

Table 6. This malware classification manner seeks training error minimization, which tailors the model towards the peculiarities of the dataset and may result in weak generalization. The trade-off between training error minimization and improved generalization is leveraged by deep feature extraction and applying conventional ML for classification. Deep features are extracted using TL based fine-tuned CNN architectures. These features are feed into SVM. This strategy showed improvement in the results.

**Table 5.** Softmax probabilistic-based implementation of customized CNN for malware classification

| | Training Scheme | | | | | | | |
| --- | --- | --- | --- | --- | --- | --- | --- | --- |
| | Training from Scratch | | | | Transfer Learning based | | | |
| **Model** | Acc % | Recall | Precision | F1-Score | Acc % | Recall | Precision | F1-Score |
| **DenseNet-201** | 96.57 | 0.9054 | 0.9080 | 0.9067 | 98.13 | 0.9411 | 0.9373 | 0.9392 |
| **Resnet-18** | 96.41 | 0.9203 | 0.9176 | 0.9189 | 98.13 | 0.9416 | 0.9374 | 0.9395 |
| **GoogleNet** | 87.20 | 0.8505 | 0.8772 | 0.8637 | 97.11 | 0.9178 | 0.9199 | 0.9189 |
| **Inception-V3** | 95.72 | 0.8941 | 0.9025 | 0.8983 | 96.36 | 0.8905 | 0.8905 | 0.8905 |
| **Xception** | 94.48 | 0.9148 | 0.9153 | 0.9150 | 96.01 | 0.8809 | 0.9025 | 0.8916 |
| **ResNet-50** | 94.86 | 0.8580 | 0.8934 | 0.8753 | 96.71 | 0.8984 | 0.8840 | 0.8911 |
| **AlexNet** | 88.38 | 0.8169 | 0.8643 | 0.8399 | 97.91 | 0.9329 | 0.9299 | 0.9314 |
| **VGG-16** | 93.41 | 0.8525 | 0.8891 | 0.8705 | 97.13 | 0.9192 | 0.9268 | 0.9230 |
| **VGG-19** | 94.97 | 0.8629 | 0.8939 | 0.8782 | 97.46 | 0.9259 | 0.9224 | 0.9241 |

**Table 6.** Deep feature extracted from customized CNN models and SVM classification

| | Training Scheme | | | | | | | |
| --- | --- | --- | --- | --- | --- | --- | --- | --- |
| | Training from Scratch | | | | Transfer Learning based | | | |
| **Model** | Acc % | Recall | Precision | F1-Score | Acc % | Recall | Precision | F1-Score |
| **DenseNet** | 97.70 | 0.9286 | 0.9286 | 0.9286 | 98.39 | 0.9483 | 0.9452 | 0.9468 |
| **ResNet18** | 98.07 | 0.9387 | 0.9368 | 0.9377 | 98.37 | 0.9463 | 0.9427 | 0.9445 |
| **GoogleNet** | 97.54 | 0.9284 | 0.9232 | 0.9258 | 97.60 | 0.9270 | 0.9250 | 0.9259 |
| **Inception** | 97.54 | 0.9262 | 0.9268 | 0.9265 | 96.95 | 0.8965 | 0.9860 | 0.9391 |
| **Xception** | 97.21 | 0.9170 | 0.9141 | 0.9156 | 98.31 | 0.9421 | 0.9467 | 0.9444 |
| **ResNet50** | 97.59 | 0.9289 | 0.9258 | 0.9273 | 97.72 | 0.9301 | 0.9284 | 0.9292 |
| **AlexNet** | 97.59 | 0.9279 | 0.9286 | 0.9282 | 98.15 | 0.9417 | 0.9378 | 0.9397 |
| **VGG16** | 97.43 | 0.9246 | 0.9221 | 0.9234 | 97.91 | 0.9355 | 0.9304 | 0.9329 |
| **VGG19** | 97.64 | 0.9281 | 0.9284 | 0.9283 | 98.23 | 0.9436 | 0.9420 | 0.9428 |

These experiments concluded that these two architectures are performing well in our customised framework.

Based upon the results from the experiments performed following conclusion can be made based upon accuracy and F1-Score.

1. DenseNet-201 and ResNet-18 > all other CNNs
2. Deep Feature Based SVM Classification > SoftMax Probabilistic-based Classification
3. TL Based Model > Training from Scratch Models

The stage is set for using DenseNet-201 and ResNet-18 through Deep Feature-based classification and TL based training scheme. One question still remains unanswered that why we selected two instead of one. This question is answered in the following section.

We also concluded from the experiments that some of the CNN models are performing well for a set of malware classes while missing others.

It is obvious from precision and recall, Figure 7 and Figure 8 respectively that GoogleNet, Inception-V3, Xception, AlexNet, VGG-16/19 misclassified Autorun.K in Yuner.A along with some other abnormalities in Wintrim and SwizorI. Two CNN architectures, ResNet-18 and DenseNet-201, are proposed for the Malware analysis. The customized ResNet-18 correctly classified Autorun.K and Yuner.A samples of the malware family. Whereas the customized DenseNet-201 performs almost similar precision as ResNet-18 model, but recall is improved. Both models improve precision while maintaining a reasonable detection rate. ResNet-18 improved the precision rate while having an equal detection rate as the existing CNN. In the Malware challenge, the major concern is improving the detection rate (Reducing False negative). Therefore, DenseNet-201 reduces a significant amount of False-negatives. ResNet-18 improve the precision rate while DenseNet-201 detection rate. Therefore, deep concatenated feature space-based MC is proposed to improve the detection rate while retaining the precision rate. ResNet-18 learned more prominently Autorun.K class while DenseNet-201 learned the rest of the malware classes. The concatenated feature space efficiently learned both classes by concatenating (ensembling) the decision of both the networks. Finally, the new proposed approach (DFS-MC) enhances the detection rate by correctly classifying Wintrim and SwizorI.

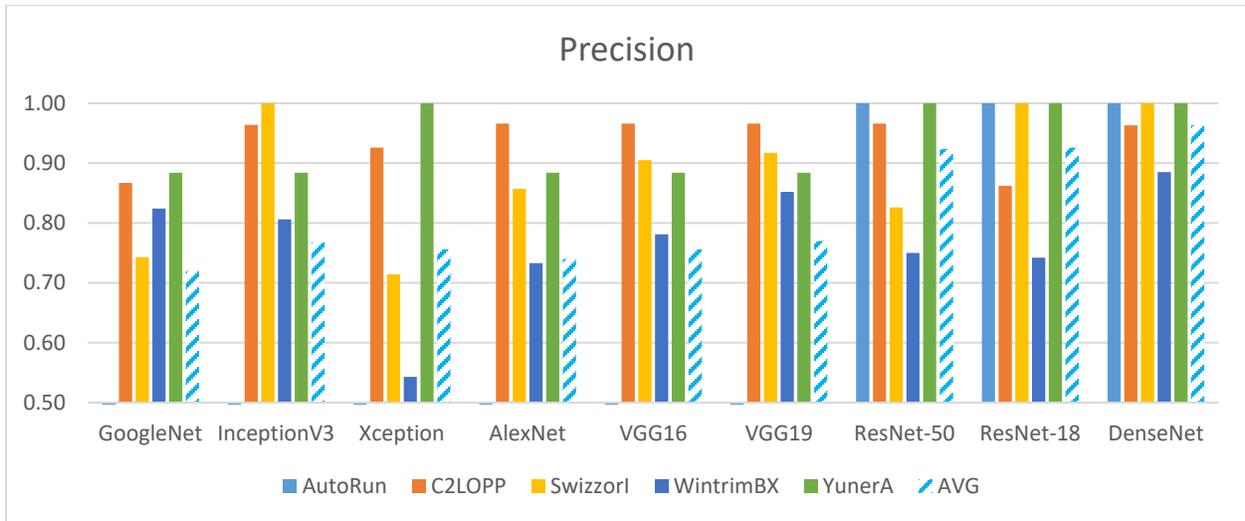

**Figure 7.** Performance (Precision) comparison of implemented CNN models

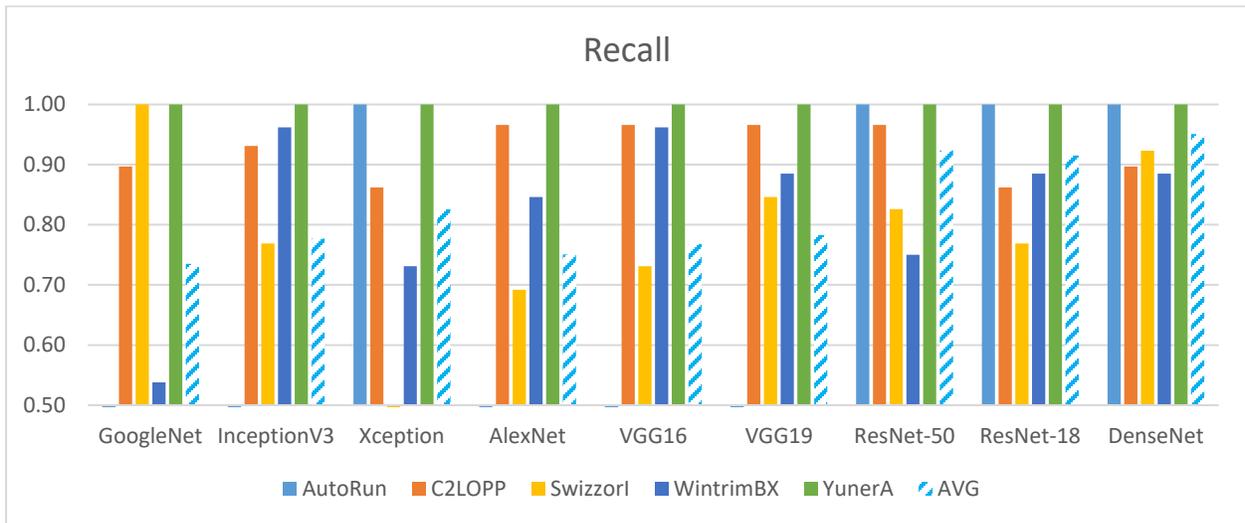

**Figure 8.** Performance (Recall) comparison of implemented CNN models

### *4.1.  Comparative Analysis with reported work*

Table 7 shows the performance of the different work in the area of malware classification. Most of the work compared here are performed using dataset portioning of 70-30 or higher. Only LGMP used dataset partitioning of 60-40 with ACC 90.23. Most of the work does not represent the F1-Score for the malware imbalanced dataset. Danish et al. [24] and Cui et al. [37],[38] showed their performance by quoting precision and recall too.

Table 7. Comparative analysis of the reported work

| Technique | %Accuracy | F-Score | Precision | Recall |
|---|---|---|---|---|
| LGMP-2018(Encoder based)++ [23] | 90.23 | - | - | - |
| LGMP-2018(Cluster based)++ [23] | 89.58 | - | - | - |
| Danish et al (70-30) [24] | 99.50 | - | 99.50 | 99.46 |
| NSGA-II∗∗ [37] | 97.60 | - | - | 88.40 |
| Cui et al. [38] | 94.5 | - | 94.6 | 94.5 |
| VGG, End-to-End [39] | 90.77 | - | - | - |
| VGG, SVM [39] | 92.29 | - | - | - |
| ResNet + SoftMax [39] | 98.62 | - | - | - |
| Natraj et al. ( 10 fold cross validation ) [13] | 98.08 | - | - | - |
| S. Lad et al. (CNN + SVM) [16] | 98.03 | - | - | - |

## 4.2. Performance of the proposed DFS-MC

All the experiments were performed at a dataset distribution of 60:40. The DenseNet-201 and ResNet-18 outperform on unseen data (40 % test data). The aim of the experiments was to surpass the performance of the existing malware classification reported framework, as shown in Table 7. Moreover, we highlighted the best customized classification Models (ResNet-18 and DenseNet-201) in our proposed malware classification framework.

The standard measure for an imbalanced dataset is F-score [40]. However, previous results were reported in terms of Accuracy. Therefore, we compare our tech with others based on Accuracy.

Table 8. Performance of DFS-MC

| Model | Accuracy % | Recall | Precision | F1-Score |
|---|---|---|---|---|
| Proposed | 98.61 | 0.9632 | 0.9627 | 0.9630 |

This improvement in the performance is because of the concatenation of hybrid deep feature space of two best performing customized models (ResNet-18 and DenseNet-201). In essence, these proposed approaches learn highly discriminative features to classify malware families. Moreover, the proposed approach (DFS-MC) achieved the highest classification performance, as shown in

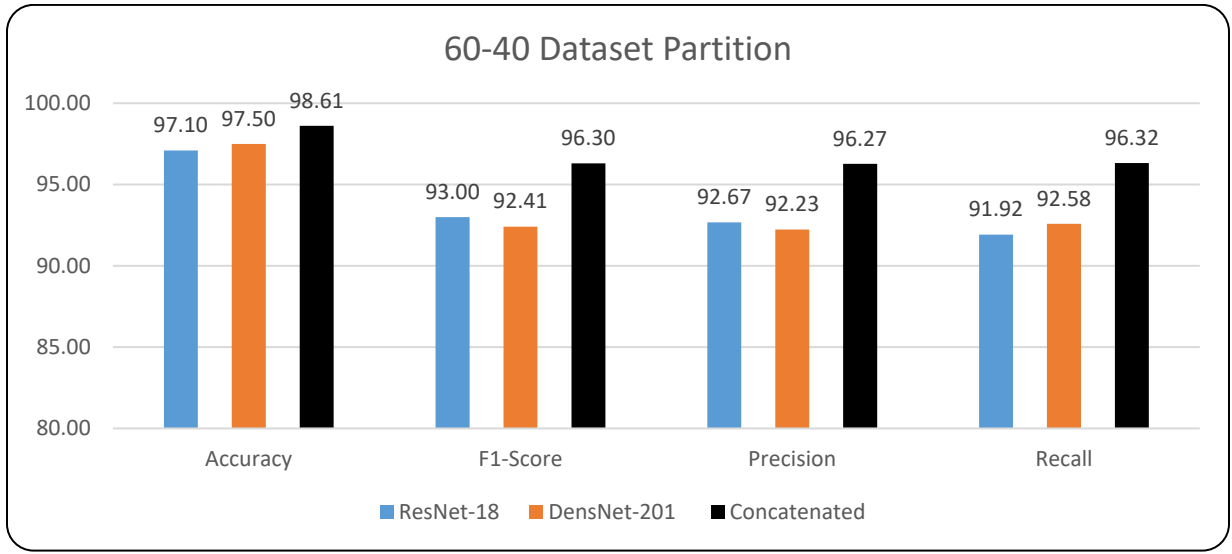

**Figure 9.** Performance improvement of the concatenated CNN

Table 8. The performance of the CNN models alone and in concatenated form, is also depicted in Figure 9.

## 5. CONCLUSION

Signature-based anti-malware products can only identify the registered malware. Trending towards the dynamic or heuristics based malware detection techniques may be resource and time-intensive. Machine learning techniques have adequate potential to handle this challenge. Traditional malware analysis technologies should be transformed into machine learning and deep learning as deep learning, specifically, can handle the huge amount of dynamics mimicked in the malware repositories. This study proposed a hybrid-learning malware classification framework that has not only identified the best performing CNN models for malware classification but also showed performance on large unseen dataset partition (60-40). Our proposed hybrid-learning framework has merged deep feature learning and conventional machine learning techniques. The identified best performing CNNs has achieved the highest Accuracy and F1-Score (Accuracy 98.61% and F1-Score 0.9630). Precision and Recall is also improved when these models are concatenated.